\newcommand\ttm[4]{\left(\begin{array}{cc}
#1 & #2 \\ #3 & #4
\end{array}\right)}
\newcommand\tom[2]{\left(\begin{array}{c}
#1 \\ #2
\end{array}\right)}
\newcommand\lC{{\mathbb C}}
\newcommand\lE{{\mathbb E}}
\newcommand\lK{{\mathbb K}}
\newcommand\lR{{\mathbb R}}
\newcommand\lV{{\mathbb V}}

\documentclass{article}
\usepackage[hypertex]{hyperref}
\usepackage{amsfonts}
\usepackage{graphicx}

\title{A toy quantum analog of enzymes}
\author{George Svetlichny\footnote{Departamento de Matem\'atica, Pontif\'{\i}cia Universidade Cat\'olica, Rio de Janeiro, Brazil \newline
svetlich@mat.puc-rio.br \hfill \url{http://www.mat.puc-rio.br/\~svetlich}}}
\begin{document}
\maketitle

\begin{abstract}
We present a quantum system incorporating qualitative aspects of enzyme action in which the possibility of quantum superposition of several conformations of the enzyme-substrate complex is investigated. We present numerical results showing quantum effects that transcend the case of a statistical mixture of conformations.
\end{abstract}

\section{Introduction}

There is by now varied empirical evidence that enzyme action is accompanied by conformation dynamics of the enzyme-substrate complex\cite{zavo-hadj:biopolymeres99.263, nageetal:PNAS108.10520, eisenetal:science295.1520, falk:science295.1480}. If several conformations are involved in catalysis a natural question is then whether the complex in such conditions is best described as a mixed state of the various possible conformations or as a pure state with the conformations in quantum superposition. In the latter case the complex is a small Schr\"odinger cat state. Once formed, environmental interactions will modify this state and one possibility is that it can ``decohere" and resemble more the mixed state, but other processes are possible.   Even under decoherence to a mixed state,  given the time scales involved, the cat state may still offer quantum advantages to the catalytic process. It is generally argued that decoherence acts so quickly that no such quantum advantage can be had. Though estimates of decoherence rates generally uphold this view, we feel that such a conclusion may be premature. The cellular environment is fairly complex, heterogeneous, crowded and compartmentalized, presenting a quantum channel whose characteristics have not been well established. By now there is a large literature (too numerous to list) on environmental influences  of a much more varied character than inhibiting certain quantum effects that would be present otherwise, but in fact acting to create, maintain, or facilitate them. See  \cite{mohsetal:JCP129.174106} for an example in photosynthesis.  Another possibility of long-lived quantum coherence in biological system is proposed in \cite{vattetal:plosone9.e98017}. Because of these considerations we have not undertaken any decoherence studies at this time. To address our main issue we present a toy analog of the enzyme process. We do not call our system a model, as any correspondence to actual enzymes would be stretching the point. The analog system however does incorporate some of the qualitative features of true enzymes.  We represent the conformations of the complex by a finite discrete quantum variable and the process of catalysis by a single real quantum variable with a quantum particle on a line representing the reaction coordinate of conventional representations of enzyme dynamics. The enzyme-substrate interaction is described by potential energy barriers in each conformation. For simplicity's sake and to facilitate computation (much of which is exact symbolic) we assume delta function potentials for these barriers. One of the quantities of interest is the transmission coefficient of the quantum scattering states of this system. This represents what would be called the ``tunneling" rate as is usually understood in  enzyme action, but is a more inclusive concept. We do not incorporate in the analog any thermal considerations. Numerical results for the case of two conformations  show  that for many possible systems superposition of conformations can enhance or suppress the transmission rate in relation to the mixed state model even if one of the barriers is non-penetrating. Beyond this, some components of the ``particle" variable can have a bound state form even though all interactions are repulsive. This cannot happen in the mixed state model. Except possibly for the ``half-bound" states, these conclusions should not be considered surprising as one expects generally, from normal quantum dynamics, that the enzyme-substrate complex would enter into a superpositions of conformations and be bound as such for a significant subset of system parameters, and again for a significant subset of system parameters, that there would be quantum effects that transcend those present in stochastic mixture of conformations. It is instructive nevertheless to see explicit confirmations of this in numerical examples. Whether analogs of such effects play any important role in real enzyme dynamics is largely dependent on the ``decohering" processes of the cellular environment. An exploration of these processes, even on a toy analog level, is beyond the present level of analysis.

In the last subsection of this paper we use the same formal system, again with just two conformations,  to present an analog not of the enzyme dynamics but possibly of the initial binding of the enzyme to the substrate by letting one or both of the delta function potentials be attractive. Numerical results show that indeed, as expected, the initial binding can be of a quantum superposition of conformations. The examples are instructive but a true quantum analog of this process needs further development.

\section{The delta potential on a line}

Consider the Shr\"odinger equation

\begin{equation}\label{Shdelta}
 i\frac{\partial}{\partial t}\Psi = -\frac12\frac{d^2}{d q^2}\psi + v\delta(q)\Psi
\end{equation}
To the right and left of \(q=0\) the equation describes a free particle and so has sinusoidal waves (pure phase) as energy eigenfunctions. In three dimensions these are called \emph{plane waves}. Writing \(\Psi(q,t)=\exp(-iEt)\psi(q)\) then one has to have
\begin{equation}\label{pwsol}
  \psi(q)= \left\{\begin{array}{cl}
  Ae^{ikq}+Be^{-ikq},& q<0\\
  Ce^{ikq}+De^{-ikq},& q>0
  \end{array},\right.
\end{equation}
with \(E=k^2\). We are for now assuming \(E>0\).

Continuity at \(q=0\) imposes \(A+B=C+D\). Call this number \(F\).  To take care of the delta potential one needs a discontinuity in \(\psi'(q)\) at \(q=0\) so that the derivative of this discontinuity gives a delta function that cancels the potential. One must then have:
\begin{equation}\label{deltacond}
  \psi'(0+)-\psi'(0-)=v\psi(0),
  \end{equation}
which translates to
\begin{equation}\label{deltacond2}
  ik(C-D-A+B)=2vF
\end{equation}

Consider now the case that a flux of particles comes in from the left, part of which is reflected by the potential and part of which tunnels through. There are no particles coming in from the right. The above wave function is also an eigenstate of the momentum operator \(\displaystyle p=\frac1i\frac{d}{d q}\) with eigenvalue \(k\). As we've set the mass to be \(1\) the velocity of the particle is \(k\). The incoming flux from the left is then \(k|A|^2\) seeing as the density of these particles is \(|A|^2\). The outgoing flux on the right is \(k|C|^2\). If we assume \(D=0\), that is there are no incoming particles from the right, then the fraction of left incoming particles that tunneled through is  \[T=\frac{|C|^2}{|A|^2}\] and is known as the \emph{transmission
coefficient.} Without loss of generality we now take \(D=0\) and \(A=1\), so \(F=1+B\). The wave function that has tunneled through is then \(Fe^{ikq}\). Condition (\ref{deltacond2}) then reads \(ikB=(1+B)v\) and we finally find:
\begin{equation}\label{F}
  F=\frac{ik}{ik-v}.
\end{equation}

With this we find
\begin{equation}\label{tc}
  T=\frac{k^2}{k^2+v^2}=\frac{E}{E+v^2}.
\end{equation}

This is the fraction of the incoming beam that has tunneled through. It tends to \(1\) as the energy increases, as to be expected.

This is all that one can say for \(v>0\), however when \(v<0\) there is also a negative energy solution which is a bound state. One must have in this case, up to a normalization constant
\begin{equation}\label{bsol}
  \psi(q)= \left\{\begin{array}{cl}
  e^{Kq},& q<0\\
 e^{-Kq},& q>0
  \end{array},\right.
\end{equation}
with \(K>0\).

The delta potential now imposes \(K+v=0\) or \(K=|v|\) and one has \(E=-v^2\) as the energy of the bound state.

\section{The enzyme analog}

Here we have one quantum variable that represents the substrate-product system and another one representing the conformation of the  enzyme-substrate complex. The first will be modeled as a particle on a line as above, the other will have a finite number of conformations and so will be a discrete variable\footnote{A discrete quantum model for conformations was introduced in \cite{cona:biosystems27.223}. Our formalism in other aspects is however rather different.} taking on the (conventional) values \(s=1,2, \dots, n\). The combined wave function will then be \(\Psi(q, s,t )\) which for convenience will be represented as a column matrix
\begin{equation}\label{Psic}
  \Psi(q, t)=\left[\begin{array}{c}
  \Psi(q,1, t)\\ \Psi(q,2,t)\\ \vdots \\ \Psi(q, n,t)
  \end{array}\right].
\end{equation}
The Hilbert space is then \(L^2(\lR)\otimes \lC^n\)  and the enzyme can be thought of as a qudit of dimension \(n\). In general the two variables are entangled.
In spite of the attention given to entanglement in the quantum information literature, what is more to the point here is the superposition of the conformation states of the enzyme-substrate complex. We assume that for each conformation of the enzyme, it subjects the other variable to a delta function energy barrier \(v_s\delta(q)\) with \(v_s>0\).

The Schroedinger equation for the combined system is now:

\begin{equation}\label{Schenz}
 i\frac{\partial}{\partial t}\Psi = -\frac12\frac{d^2}{d q^2}\psi + \delta(q)\lV\Psi+M\Psi,
\end{equation}
where \(\lV = \hbox{diag}(v_1,v_2,\dots,v_n)\)  is the diagonal matrix of the strengths of the delta potentials and M is a \(n\times n\) hermitian matrix describing the quantum dynamics of the enzyme-substrate conformations.

In this expression the quantity \(H_I= \delta(q)\lV\) is the \emph{interaction Hamiltonian,} and \(H_0-\frac{d^2}{d q^2}+M\) is the \emph{free Hamiltonian.} The basis in which (\ref{Schenz}) is written is one in which \(H_I\) is (block) diagonal, and will be referred to as the \emph{interaction basis.}

Unless \(M\) is diagonal, then equation (\ref{Schenz}) cannot be solved by pure phase solutions on either side of the \(q=0\) barrier, but a unitary transform of it can be. As \(M\) is hermitian, there is a unitary \(U\) such that \(UMU^*=\lE=\hbox{diag}(e_1,e_2,\dots,e_n)\) which are the energy eigenvalues of the enzyme alone. Letting  now \(\tilde\Psi=U\Psi\)
equation (\ref{Schenz}) becomes
\begin{equation}\label{Schenzt}
 i\frac{\partial}{\partial t}\tilde\Psi = -\frac12\frac{d^2}{d q^2}\tilde\Psi + \delta(q)U\lV U^*\tilde\Psi+\lE\tilde\Psi,
\end{equation}

Equation (\ref{Schenzt}) is written in a basis in which the free Hamiltonian is (block) diagonal and will be called the \emph{free basis}. Non-trivial quantum effects can only happen if the interaction and the free basis are misaligned for otherwise the system of equations separates into independent ones.

On either side of the barrier the various components of \(\tilde\Psi\) decouple and satisfy a free particle Schroedinger equation. So we can take, assuming stationary waves now, that \(\tilde \Psi_s(q,t)=\exp(-iEt)\tilde \psi_s(q)\) where
the equation that \(\tilde\psi(q)\) must satisfy is:
\begin{equation}\label{Schenzte}
E\tilde\psi = -\frac12\frac{d^2}{d q^2}\tilde\psi + \delta(q)U\lV U^*\tilde\psi+\lE\tilde\psi,
\end{equation}
from which we deduce that for each \(s\) we have phase waves of the form
\begin{equation}\label{pwsols}
 \tilde \psi_s(q)= \left\{\begin{array}{cl}
  A_se^{ik_sq}+B_se^{-ik_sq},& q<0\\
  C_se^{ik_sq}+D_se^{-ik_sq},& q>0
  \end{array},\right.
\end{equation}
with
\begin{equation}\label{E}
E=k_s^2+e_s
\end{equation}
fixing thus the value of \(k_s\) for each value of the energy.

Continuity at \(q=0\) entails \(A_s+B_s=C_s+D_s\). Call this common value \(F_s\). The delta function potentials now impose the following condition on the discontinuity of \(\psi'(q)\) at \(q=0\).

\begin{equation}\label{delcondtild}
  \tilde\psi'(0+)-\tilde\psi'(0-)=2U\lV U^*\tilde\psi(0),
  \end{equation}

For tunneling solutions we set \(D_s=0\) so one has \(C_s=A_s+B_s=F_s\) and we impose \(\sum_sk_s|A_s|^2=1\) to normalize the flux arriving from the left. The transmission rate is then \(T=\sum_sk_s|F_s|^2\).  In this case we have:
\begin{eqnarray}\label{enzydat}
  \tilde\psi_s(0) &=& A_s+B_s \\
   \tilde\psi_s'(0+)-\tilde\psi_s'(0-) &=& 2ik_sB_s.
\end{eqnarray}

Let \(\lK=\hbox{diag}(k_1,k_2,\dots,k_n)\) and arrange the \(A_s\), the \(B_s\) and the \(F_s\) into column vectors \(\bf A\), \(\bf B\), and \(\bf F\) respectively. We now have
\(i\lK{\bf B}=U\lV U^*({\bf A+B})\). For fixed \(\bf A\) this can be solved for \(\bf B\) which can then be used to calculate the transmission coefficient.
We have \( \bf B=(i\lK-U\lV U^*)^{-1}U\lV U^*{\bf A}\) and after a little manipulation we have (compare to (\ref{F}))
\begin{equation}\label{fsol}
  {\bf F}=i(i\lK-U\lV U^*)^{-1}\lK{\bf A}.
\end{equation}

For \(E\) greater than the largest \(e_s\) one can take  \(k_s>0\), hence real, and seeing that \(U\lV U^*\) is hermitian, the inverse matrix in the formula above exists, and we have a well defined scattering solution. Possible solutions for lower values of \(E\) need a more detailed analysis.

\subsection{Quantum Advantage}\label{qad}
The principal question addressed in this text is whether the possibility of superposition of the conformations provides any advantage in relation to the case of a mixed state of separate conformations. By advantage we mean either enhancement or suppression of transmission rates, or the presence of new quantum effects that cannot exist in the mixed state. We consider suppression beyond what happens in the mixed state as an advantage, since it may be the case that in real enzymes the transition in some conformation of the substrate to some product other that the wanted one may be undesirable.  Among the possible systems with superposition one can also consider other comparison criteria and we do so in specific examples below.

Now for a fixed conformation \(s\), the transmission rate is given by
\begin{equation}\label{tce}
 T_s= \frac{k_s^2}{k_s^2+v_s^2}.
\end{equation}
If now one assumes that in the intermediate state one has a mixture and not a superposition of  conformations of the enzyme-substrate complex, then the transmission rate would be somewhere in between these two numbers, so if it is outside then it signals advantage and is a sure sign of superposition.

 Under the superposition hypothesis the comparison with these two numbers in not necessarily the most relevant one. In equation (\ref{Schenzt}), one sees that each component would be subject to interaction with several delta potentials, thus one should actually consider the transmission rates of each wave number through each potential, that is:

\begin{equation}\label{tceall}
 T_{ss'}= \frac{k_s^2}{k_s^2+v_{s'}^2}
\end{equation}

Numerical results on the two-dimensional case treated below suggest that the transmission rate is always between the smallest and the largest of these. This seems entirely reasonable and is probably a theorem, but has not yet been established.

Another, and for some purposes maybe better, comparison would be between the present system and ones with aligned interaction and free bases. In these the superpositions would be of independent systems, so the advantage would be due merely to superposition.  To this end we consider variations on (\ref{Schenzt}) of the form
\begin{equation}\label{Schenztpi}
 i\frac{\partial}{\partial t}\tilde\Psi = -\frac12\frac{d^2}{d q^2}\tilde\Psi + \delta(q)\lV_\pi \tilde\Psi+\lE\tilde\Psi,
\end{equation}
where \(\lV_\pi\) is the diagonal matrix obtained from \(\lV\) by permuting the diagonal entries by a permutation \(\pi\). This system has the same energetic profile as (\ref{Schenz}) in terms of the interaction strengths of the potentials and the energetic eigenvalues of the conformation complex, but no mathematical coupling between the components. Comparison of the transmission rate through these systems (for all permutations) and that of (\ref{Schenzt}) gives another feel for ``quantum advantage".

In the two-dimensional case, we'll present all these comparisons, along with some others.

Among the cases of aligned interaction and free bases, is the degenerate one, that in which the matrix \(\lE\) has repeated entries. One expects that as degeneracy increases quantum advantage would decrease and with \(\lE\) proportional to the identity there is no advantage in relation to the mixed state situation. This is born out in the two-dimensional examples.

\section{The \protect\(2\times 2\protect\) case}

\subsection{Enhanced and suppressed transmission} \label{est}

We take now \(n=2\) the simplest case and, as above, \(v_s>0\), that is, for fixed conformations we have a potential barrier. Now \(U\lV U^*\) is a hermitian matrix with spectrum \(\{v_1,v_2\}\). This has trace \(v_1+v_2\) and determinant \(v_1v_2\). We assume, with no loss of generality, \(v_1> v_2\). The most general matrix with these properties can be given by:
\begin{equation}\label{vm}
  \left(\begin{array}{cc}
  v_1-y(v_1-v_2) & re^{i\theta}\\
  re^{-i\theta}& v_2+y(v_1-v_2)
  \end{array}\right)
\end{equation}
where the trace condition is already satisfied and
 to satisfy the determinant condition one must have \(0\le y\le 1\) and \(r=\sqrt{y(1-y)}(v_1-v_2)\). The phase \(\theta\) is arbitrary.

Let now \(0\le h\le 1\)  and set
\begin{eqnarray}\label{a1}
 A_1 &=& \frac h{\sqrt{k_1}} \\ \label{a2}
 A_2 &=& \sqrt{\frac{1-h^2} {k_2}}e^{i\phi}.
 \end{eqnarray}
 This provides, without loss of generality, all the possible values for \(\bf A\) with normalized flux. The problem is now completely parameterized by the eight variables \(v_1,v_2,k_1,k_2, y, \theta,h\) and \(\phi\).

  We are now of course interested in \(T=k_1|F_1|^2+k_2|F_2|^2\) and compare this to the values of the transmission coefficients in comparison systems as discussed in Subsection \ref{qad}.

 A full algebraic analysis has not been attempted yet. Some numerical results however are interesting. For the values
  \[v_1=18, v_2=10;\,
 k_1=20, k_2=15;\,
 h=3/4, \phi=\pi/4,\]
 one computes\footnote{All floating point numerical results will be displayed to four decimal places.}
\begin{center}
\(\begin{array}{ll}
T_{11} = 0.5525,& T_{12} = 0.8000,\\
T_{21} = 0.4098, &
T_{22} = 0.6923.
\end{array}\)
\end{center}

The figure that follows now is the plot of \(T\) as a function of \(y\) and \(\theta\). These thus represent possible \emph{systems}, that is conformation dynamics. We are interesting in seeing if there are possible systems that present certain behavior. The planes show the values of the four transmission rates \(T_{ij}\). The two red planes correspond to wave number \(k_1\) and the two yellow ones to \(k_2\).  The higher of the two correspond to the lower potential, \(v_2\),  and the lower ones to the higher potential, \(v_1\). One sees that portions of the surface are both above and below the two middle planes. Thus superposition can both enhance and suppress tunneling in comparison to a mixed intermediate state. The whole surface falls between the highest and the lowest plane as is expected.

\begin{center}
\includegraphics[scale = 0.3]{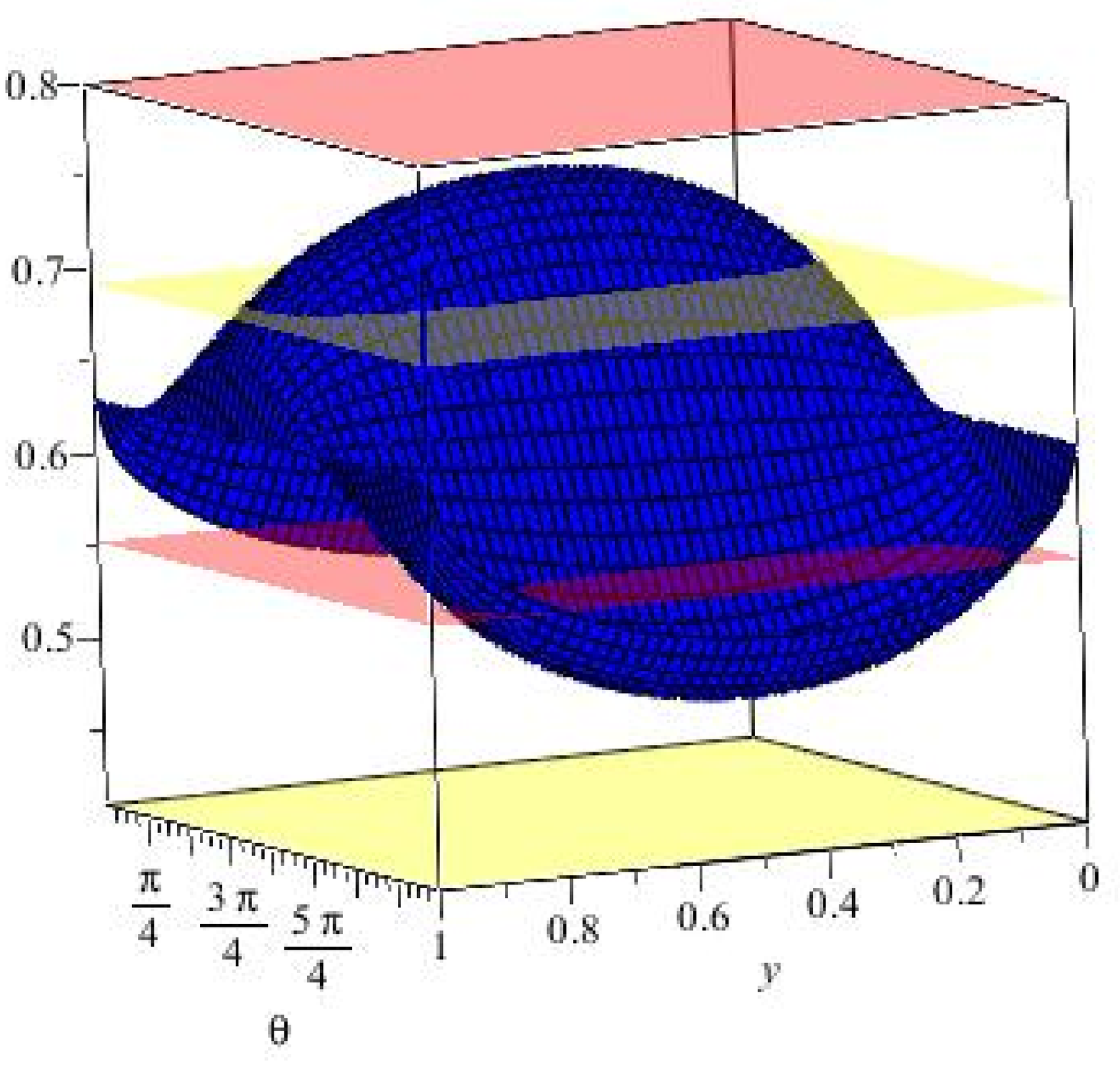}
\end{center}

For the particular values of \(y=0.5\) and \(\theta=3\pi/4\) one has \(T=0.7539\) about an 8.9\% increase over the highest transmission rate for the mixed state. Likewise for \(y=0.4\) and \(\theta=5.5\) one has \(T=0.2108\) which is  about 88.6\% of the lowest transmission rate for the mixed state.

The following figure compares the transmission rate to the rates obtained when there is no misalignment between the interaction basis and the free basis which happens when \(y\) is \(0\) (lower green plane) and \(1\) (upper green plane).  One sees again both quantum enhancement and quantum suppression, of considerable magnitudes in this case.

\begin{center}
\includegraphics[scale = 0.3]{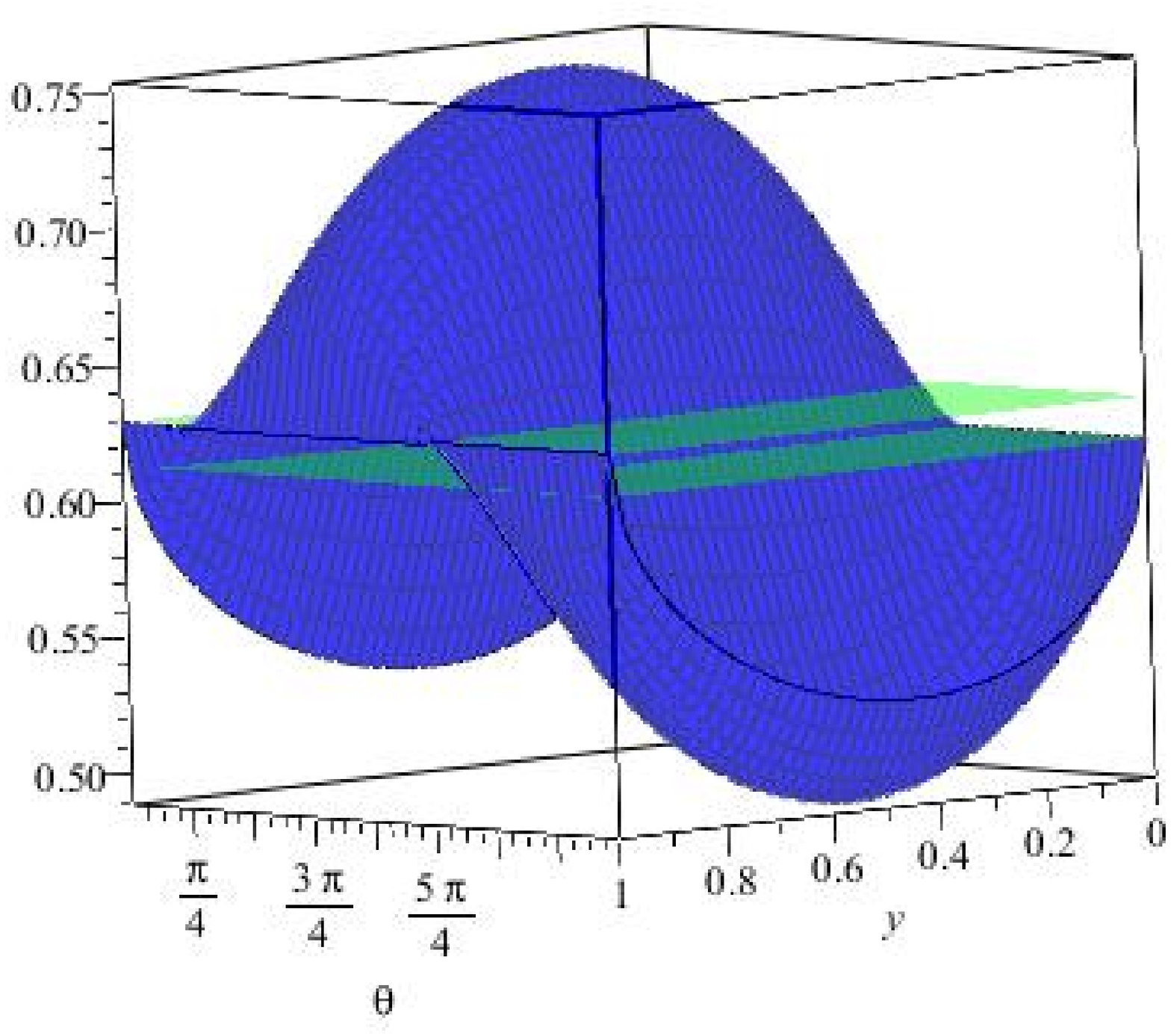}
\end{center}

As a final note, if one wants to get back to equation (\ref{Shdelta}) one has to diagonalize matrix (\ref{vm}) which can be done by some unitary \(V\) which will be unique up to diagonal unitaries which does not change the physics. One then has \(M=V\lE V^*\). One thus needs the values of \(e_1\) and \(e_2\). Now one has \(k_1^2-k_2^2=e_2-e_1\) (see (\ref{E})).  As non-relativistic physics only sees energy diferences we can choose any two values that satisfy this without changing physics. Thus the above analysis in terms of scattering waves does contain all the physics of this problem.

\subsection{Superposition induced binding}

We now address the question of whether, with potentials \(v_1\) and \(v_2\) still positive, superposition can create something like a bound state. By this we mean whether one or both components of the wave function \(\tilde \psi\) can have a form proportional to the one shown in (\ref{bsol}). One would expect, given the repulsive nature of the potentials that this would be impossible, and in fact both components of the wave-function cannot have this shape, but curiously enough one can. We start by deriving the condition for both components having exponential decaying forms for any values of the \(v_i\).

Suppose that for \(s=1,2\) one has:

\begin{equation}\label{2dogs}
 \tilde\psi_s(q)=\left\{\begin{array}{ll}
 A_se^{K_sq},& q<0\\
 A_se^{-K_sq},& q>0
 \end{array}\right.
\end{equation}
where \(K_s>0\). Let \(\tilde\lK=\hbox{diag}(K_1,K_2)\) then performing an analysis similar to before, to accommodate the delta potential, one must have:
\begin{equation}\label{2mdogs}
  (\tilde\lK+U\lV U^*){\bf A}=0.
\end{equation}
As this is a homogeneous equation, for it to have nonzero solutions one must have \(\det(2\tilde\lK+U\lV U^*)=0\). the vanishing of this determinant is now given by:
\begin{equation}\label{detb}
(K_1-K_2)(v_1-v_2)y+(K_1+v_1)(K_2+v_2)=0.
\end{equation}

 Assume now \(v_i>0\), with \(v_1>v_2\) as before, and that all variables in the matrix are fixed except for \(y\), with \(r\) expressed as a function of \(y\) as before (text following (\ref{vm})), then the determinant vanishes at the value:
\begin{equation}\label{yforb}
 y=-\frac{ (K_1+v_1)(K_2+v_2)}{(K_1-K_2)(v_1-v_2)}
\end{equation}
and what one has to check is whether this value can ever satisfy \(0\le y\le 1\).  All expressed quantities are positive. Now for \(y\) to be non-negative one has to have \(K_2>K_1\). Write \(K_2=(1+\alpha) K_1\) with \(\alpha >0\). One now has:
\[(v_1-v_2)y=\frac{(\alpha+1) K_1}{\alpha} +\frac1\alpha v_2+\frac{(\alpha+1)}{\alpha} v_1+ \frac{v_1 v_2}{\alpha K_1}.\]
From the third term on the right one sees that \(y> 1\) and so cannot lie in the required interval. Thus there are no true bound states. We shall study true bound states when one or both of the \(v_i\) is negative in subsection \ref{tb}.

Consider now the possibility that the first component of \(\tilde \psi\) has an exponentially decaying form and the second a scattering form. We shall from the start assume that there are no incoming waves from the right in the second component and thus the solution we are seeking has the following form where we've already imposed continuity of \(\tilde \psi\) at \(q=0\):

\begin{eqnarray}\label{1bc1}
  \tilde \psi_1(q) &=& \left\{\begin{array}{ll}
 Ae^{Kq},& q<0\\
 Ae^{-Kq},& q>0
 \end{array}\right. \\ \label{1bc2}
\tilde \psi_2(q) &=& \left\{\begin{array}{cl}
  Re^{ikq}+Se^{-ikq},& q<0\\
  (R+S)e^{ikq},& q>0
  \end{array},\right.
\end{eqnarray}

There is now the question of how to normalize such a wave function, as it has both a bound and a scattering component. Bound states are usually normalized to have norm one, and scattering ones to have flux or density one. We've chosen to normalize to have flux one incoming from the left by choosing \(R=1/\sqrt k\) although normalizing by \(\|\tilde\psi_1\|=1\) is another natural choice, and still others may also be useful.

Let \(\hat\lK=\hbox{diag}(-K,ik)\) then to accommodate the delta potentials one must have:
\begin{equation}\label{1b2s}
  \hat\lK\tom{A}{S}=UVU^*\tom{A}{\frac{1}{\sqrt{k}}+S}
\end{equation}
This is  solved for \(A\) and \(S\) as:
\begin{equation}\label{as}
 \tom AS= (\hat\lK-UVU^*)^{-1}UVU^*\tom 0{\frac1{\sqrt k}}.
\end{equation}
The transmission coefficient for the scattering component will be given by \linebreak \(k|\frac1{\sqrt{k}}+S|^2\) and the \(L^2\) norm of the first component will be \(\displaystyle \frac{|A|}{\sqrt{K}}\). It turns out that these quantities do not depend on the angle \(\theta\). We've not analyzed the exact solution but numerical analysis shows that such wave-functions do exist. For the values
\[ v_1=30, v_2=2;\, K=5, k=20\]
the next two figures show, as a function of \(y\), the transmission coefficient (in blue) of the second component and the norm (in red) of the first component of the wave-function.

\begin{center}
\includegraphics[scale = 0.3]{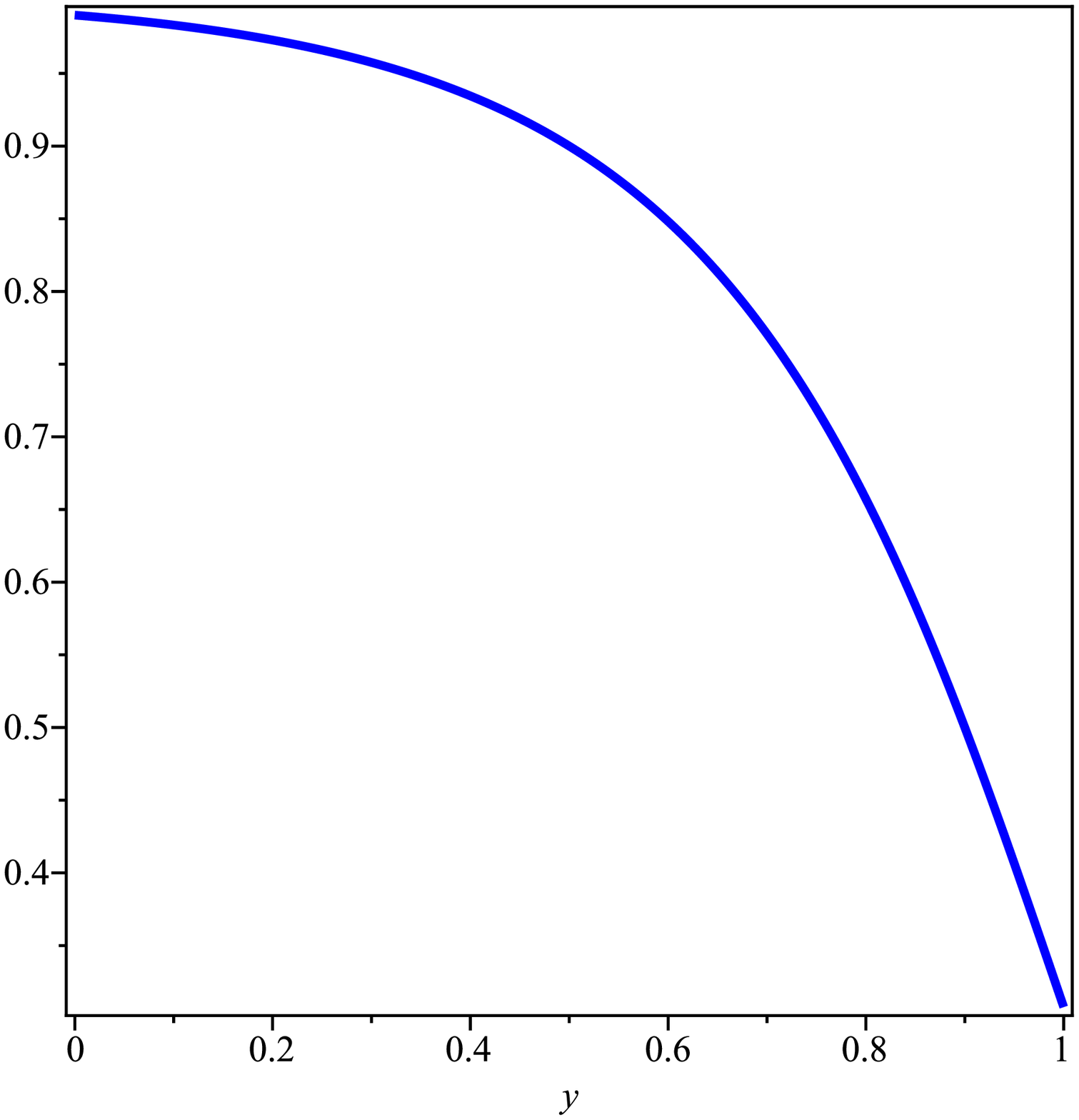} \includegraphics[scale = 0.3]{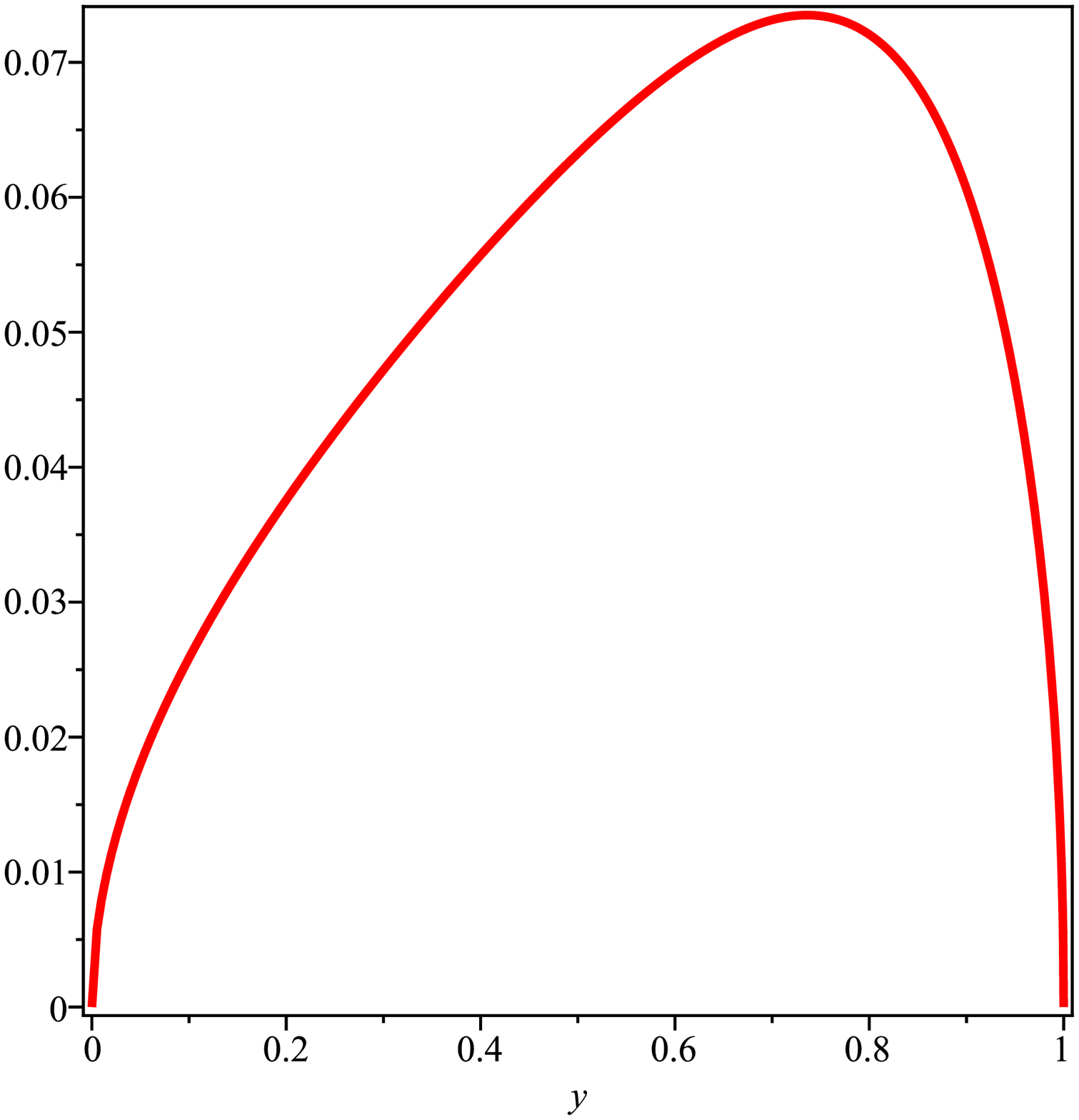}
\end{center}

The transmission coefficient of the second component interpolates between the values of fixed conformations coefficients associated to the two values of \(v_s\). These are \(0.9901\)  and \(0.3077\).
In terms of the energy eigenvalues \(e_1,\, e_2\) of the enzyme one has \(E=-K^2+e_1=k^2+e_2\) so \(e_2<E<e_1\). The red curve reaches zero at both extremes of \(y\) illustrating the fact that the bound component cannot exist without misalignment of the interaction and free bases. In this case we see no quantum enhancement nor suppression of the transmission coefficient, but the appearance of a bound state form for one of the components should be considered as a type of quantum advantage.
Now since we always assume \(e_1\neq e_2\) for otherwise \(M\) would be diagonal and the system trivial, it is always the case that such  ``half bound" states are present  and occupy the spectral interval between \(e_1\) and \(e_2\).

One also has solutions in which the second component is exponentially decreasing and the first is scattering. The plots are superficially similar to the ones shown above. In this case \(e_1<E<e_2\).

The existence of these ``half-bound" states is due to energetic contributions from the conformation dynamics which mitigate the repulsive nature of the delta potentials. Though initially these states may seem strange, in retrospect they are understandable.

What such a behaviour might mean for real enzymes is that some of the variables of the substrate may get localized and ``frozen" while processing goes on the rest, until the final unbinding of the product. And this in spite of overall repulsive nature of the conformation states.

\subsection{Infinite barrier in one channel}\label{infbar}

We now consider also the case of an infinite barrier at \(q=0\) in the first component of the $2\times 2$ case of equation (\ref{Schenz}). We symbolize such a barrier by a ``potential" \(V_b(q)\). One can consider two types of infinite barriers, one of which can be called the \emph{infinite wall} in which quantum particles can exist on both sides of the barrier but cannot tunnel through from one side to the other. Such a barrier corresponds  in some sense to the limit \[V_b(q) =\lim_{v\to \infty}v\delta(q)\] in the models of the previous subsections. This limit is of course symbolic to be interpreted appropriately in what follow. The other type of barrier can be called the \emph{infinite cliff} in which no particles can exist on one of the sides, \(q>0\) say,
and would correspond to:
\begin{equation}\label{cliff}
V_b(q)=\left\{\begin{array}{cl} 0 &q<0\\ \infty&q>0\end{array}\right.,
\end{equation}
which again would correspond in some sense to the limit
\begin{equation}\label{step}
V_b(q)=\lim_{V\to \infty}\left\{\begin{array}{cl} 0 &q<0\\V &q>0\end{array}\right.,
\end{equation}

The Schr\"odinger equation is now:
\begin{equation}\label{Schinfb}
 i\frac{\partial}{\partial t}\Psi = -\frac12\frac{d^2}{d q^2}\psi + \lV(q)\Psi+M\Psi,
\end{equation}
Where
\begin{equation}\label{vmb}
  \lV(q)=\ttm{V_b(q)}00{v\delta(q)}.
\end{equation}

The sinusoidal solutions of the free Schroedinger equation with any one of the  barrier types are \(e^{ikq}-e^{-ikq}\) on the side(s) of the barrier that allows particles. The transmission coefficient across the barrier is zero. For the infinite cliff the energy is \(E=k^2\) which is also the case for the infinite wall in which case the wave number \(k\) will be the same on both sides.

We shall consider here only the infinite wall case, for which diagonalization of \(M\) is still an effective step to construct energy eigenstates. For the infinite cliff this is no longer so, except in the  degenerate case, which is not interesting.

We now write the  Schr\"odinger equation as:

\begin{equation}\label{Schinf}
 i\frac{\partial}{\partial t}\Psi = -\frac12\frac{d^2}{d q^2}\psi + V_b(q)P_1\Psi+v\delta(q)P_2\Psi+M\Psi,
\end{equation}
where
\begin{equation}\label{P12}
P_1=\ttm1000,\quad P_2=\ttm0001
\end{equation}
are two orthogonal projections with \(P_1+P_2=I\).

As before let  \(U\) be a unitary such that \(UMU^*=\lE=\hbox{diag}(e_1,e_2)\), and let \(\tilde \Psi=U\Psi\). Then we have

\begin{equation}\label{Schinft}
 i\frac{\partial}{\partial t}\tilde\Psi = -\frac12\frac{d^2}{d q^2}\tilde\psi + V_b(q)Q_1\tilde\Psi+v\delta(q)Q_2\tilde\Psi+\lE\tilde\Psi,
\end{equation}

where \(Q_i=UP_iU^*\) with, as before, \(Q_1+Q_2=I\).

As \(Q_1\) is a rank one orthogonal projection, its most general form is
\begin{equation}\label{Q1}
  Q_1=\ttm\alpha{\sqrt{\alpha(1-\alpha)}e^{i\theta}}{\sqrt{\alpha(1-\alpha)}e^{-i\theta}}{1-\alpha}
\end{equation}
with \(0\le \alpha\le 1\), and \(Q_2=I-Q_1\).

As before we seek energy eigenstates \(\tilde \Psi_s(q,t)=\exp(-iEt)\tilde \psi_s(q)\) hence sinusoidal solutions of the form (\ref{pwsols}) whenever possible.

One must impose the following conditions on the components  \(\psi\):
\begin{enumerate}
 \item Because of the infinite barrier one must have \(\psi_1(0)=P_1\psi(0)=0\).
 \item To accommodate the delta potential one must have \(\psi_2'(0+)-\psi_2'(0-)  =2v\psi_2(0)\) that is, \(P_2(\psi'(0+)-\psi'(0-))=2vP_2\psi(0\).
 \end{enumerate}

 After changing base by \(U\) these conditions translate to:
 \begin{eqnarray}\label{binfc}
   Q_1\tilde \psi(0) &=&0 \\ \label{dindelc}
  Q_2 [\tilde\psi'(0+)-\tilde\psi'(0-)]&=&2 vQ_2\tilde\psi(0)
 \end{eqnarray}

Using the same definitions as before (with \({\bf A}\) given by (\ref{a1},\ref{a2})) we must have
\begin{eqnarray}\label{q1c}
  Q_1({\bf A}+{\bf B}) &=& 0 \\ \label{q2c}
i Q_2\lK {\bf B} &=& vQ_2({\bf A}+{\bf B}).
\end{eqnarray}
Using the fact that \(Q_2+Q_1=I\) and equation (\ref{q1c}), one can substitute \(I\) for \(Q_2\) in the right-hand side of (\ref{q2c}) and solve for \({\bf B}\) as a function of \({\bf A}\) to get:
\begin{equation}\label{bfroma}
{\bf B}=v(iQ_2\lK -vI)^{-1}{\bf A}.
\end{equation}
With a given \({\bf A}\),  and \({\bf B}\) calculated in this way, equation (\ref{q1c}) is automatically satisfied. To see this rewrite (\ref{bfroma}) as
\(iQ_2\lK{\bf B}=v({\bf A}+{\bf B})\). Apply \(Q_1\) to both sides to get (\ref{q1c}).

As before \({\bf F}={\bf A}+{\bf B}\) and the transmission coefficient is given by \(T=k_1|F_1|^2+k_2|F_2|^2\).

No further restrictions are necessary. In this case there are two non-zero transmission coefficients for comparison:
\begin{equation}\label{tciv}
  T_i= \frac{k_i^2}{k_i^2+v^2}
\end{equation}

As an example of the infinite wall we choose:
\begin{equation}\label{ibgew}
 v=6,\,k_1=13,\, k_2=5,\, h=1/2,\, \phi=\pi /2.
\end{equation}
The two transmission coefficients are: \(T_1=0.8244\) and \(T_2=0.4098\) which in the following figure correspond to the red and yellow planes respectively.
\begin{center}
\includegraphics[scale = 0.3]{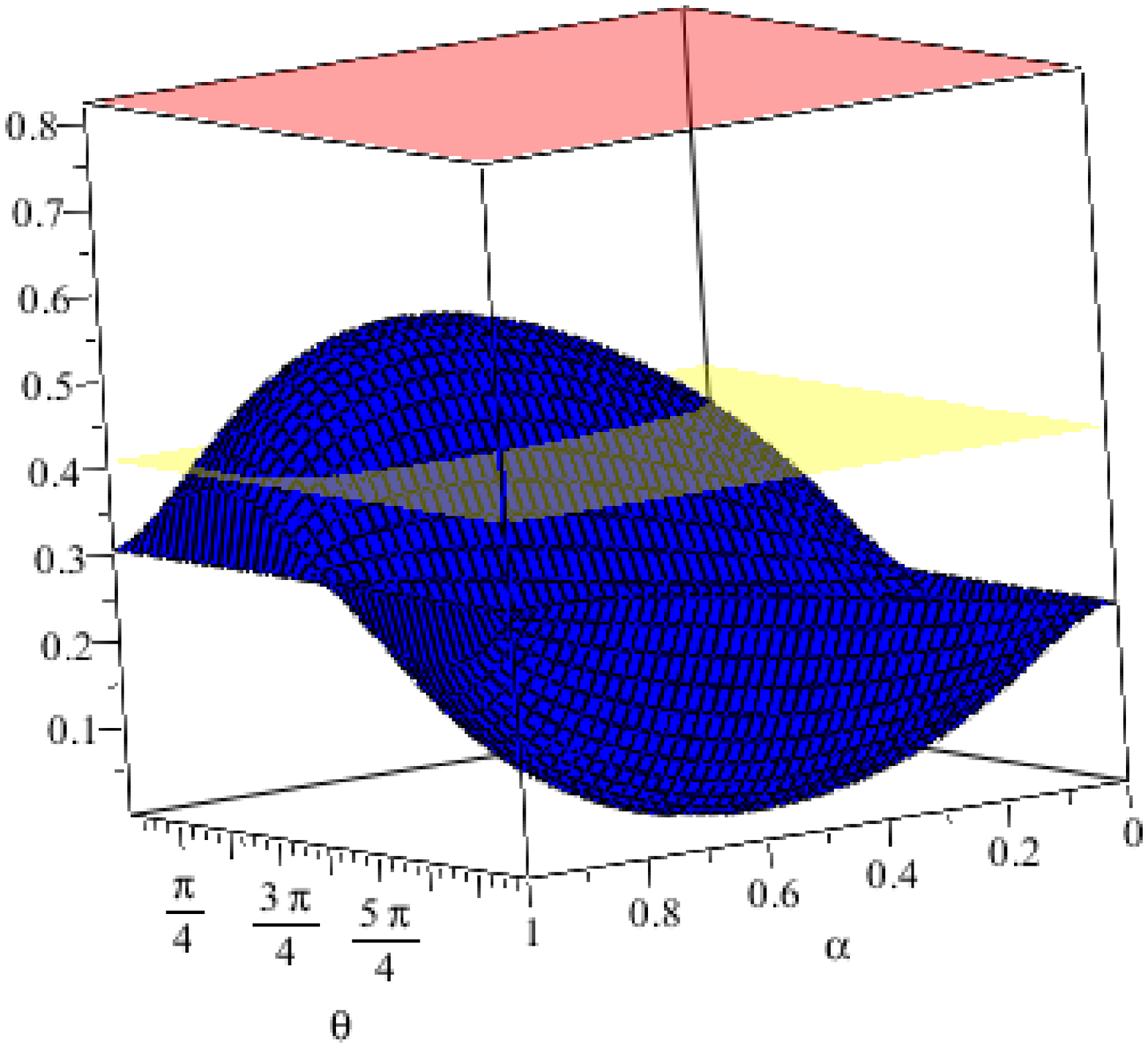}
\end{center}
If the  state were a mixture, the transmission rate would be between zero and the yellow plane. Thus there is a region of quantum enhancement due to superposition. There cannot be any suppression as the lowest transmission rate is the zero rate through the infinite barrier, and one cannot go below zero.

The advantage due to mismatch of the interaction basis and the free basis is illustrated by the following figure:
\begin{center}
\includegraphics[scale = 0.3]{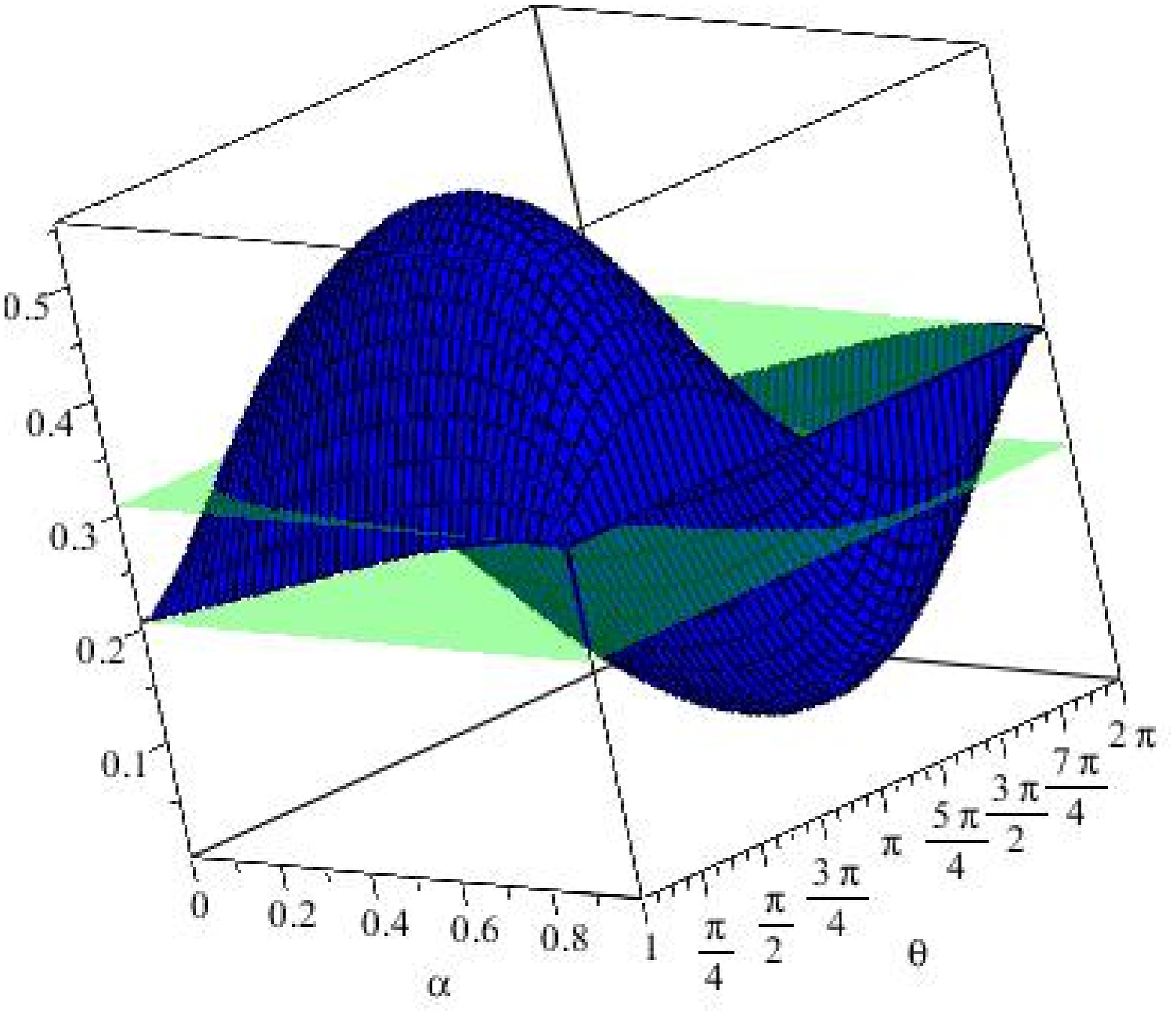}
\end{center}
where the aligned bases correspond to \(\alpha\) being \(0\) and \(1\). In this comparison there is both enhancement and suppression.

\subsection{The State View}

The transmission coefficient \(T\) depends on eight parameters: \(v_1,v_2, y, \theta, k_1,k_2, h\) and \( \phi\). Of these \(v_1,v_2, y\) and \(\theta\) are parameters solely of the system, and \(h\) and \( \phi\) are solely of the state. Because \(k_1^2-k_2^2=e_2-e_1\) the pair \(k_1\) and \( k_2\) carry information both of the system and the state. In the examples above, we've kept all state parameters and some system parameters fixed and displayed the transmission coefficients as a function of two of the system parameters \(y\) and \(\theta\) with the intent of seeing if any systems can exhibit quantum advantage according to some comparison criteria. A systematic analysis of the dependence of \(T\) on all eight parameters has not been done. In this subsection we just display a few numerical results of the complementary view, of keeping the system parameters fixed and exhibiting the behaviour upon changing the state parameters \(h\) and \( \phi\).
We chose:
\[v_1=18,\, v_2=5,\, k_1=6,\, k_2= 3,\, y=\frac12,\, \theta=\pi\]

The following figure shows the transmission coefficient as a function of \(h\) and \(\phi\) which determine the coefficients of the two components of the state. The four planes (red and yellow) have the same meaning as in Subsection \ref{est}
\begin{center}
\includegraphics[scale = 0.3]{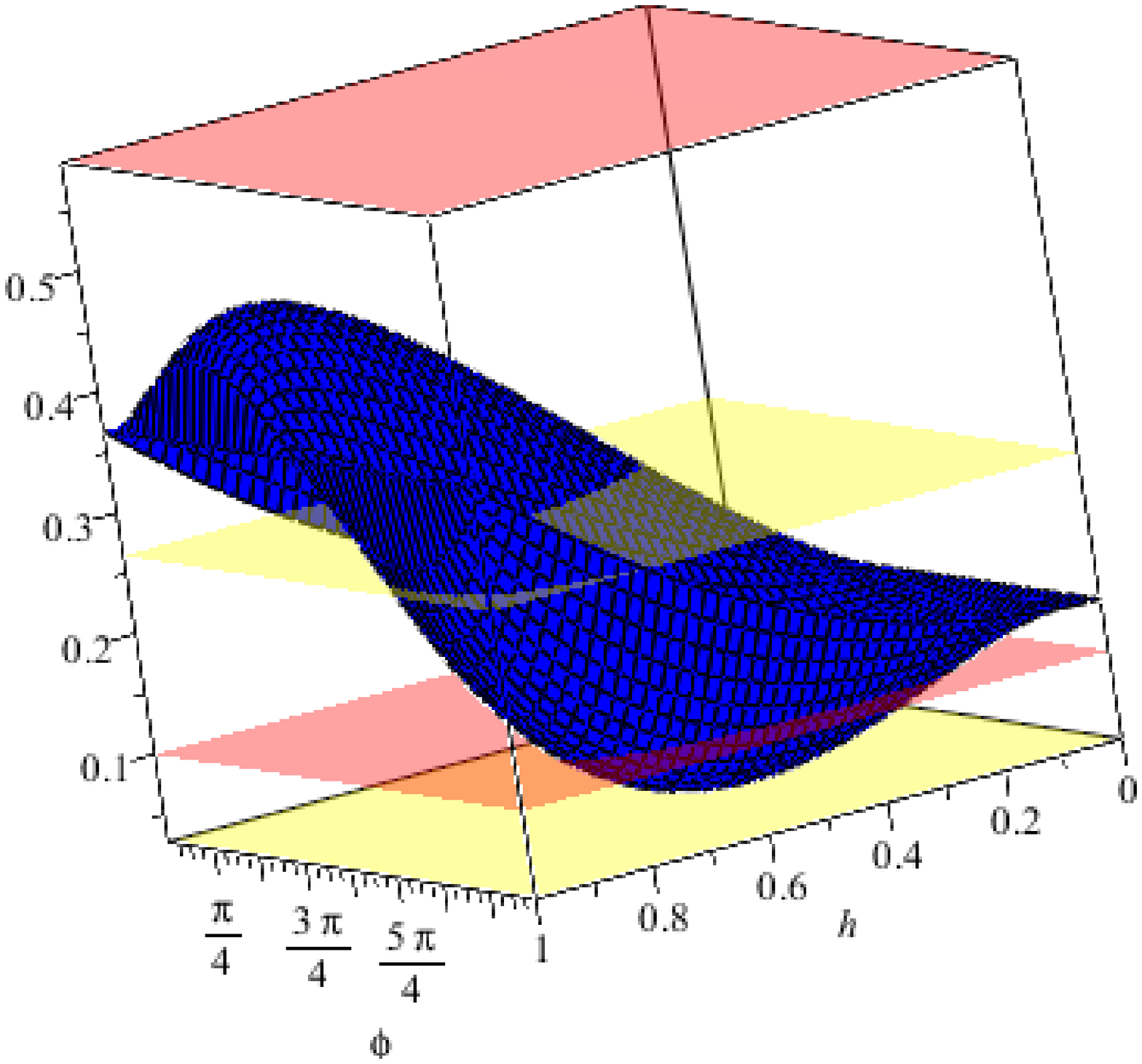}
\end{center}

One cannot in this case make a comparison with the system having other bases, as in the previous subsections. A reasonable comparison now is to the state in which one or the other component is zero, that is, an advantage in having superposition of components and hence of conformations. The following figure show the transmission coefficient with planes set at the values for \(h\) being zero or one. These values obviously do not vary with \(\phi\).

\begin{center}
\includegraphics[scale = 0.3]{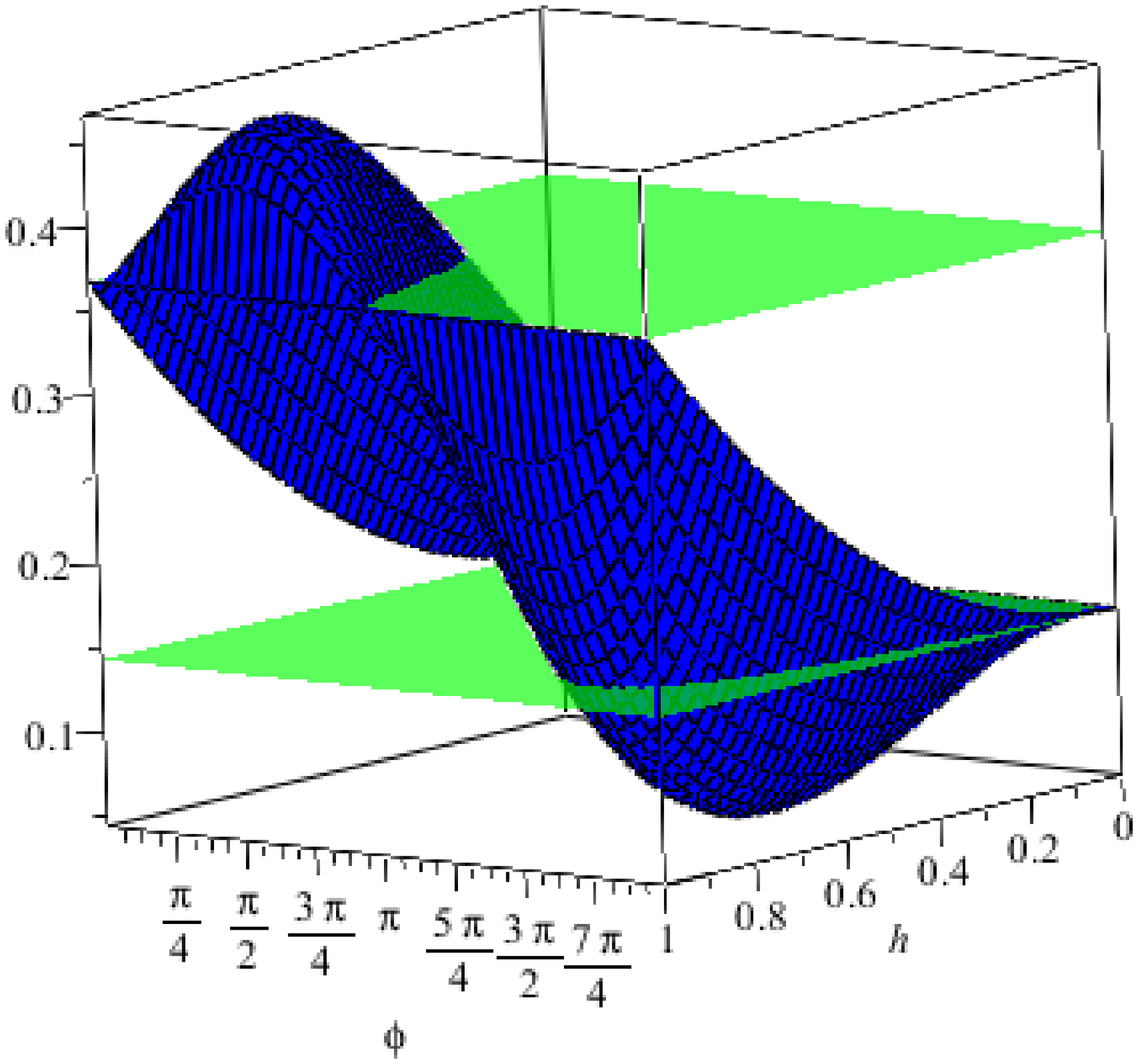}
\end{center}

\subsection{Enhanced passage through infinite wall}

In \cite{schretal:science332.1300} the authors state that  ``quantum-mechanical tunneling can supersede traditional kinetic control and direct a reaction
exclusively to a product whose reaction path has a higher barrier."  Classically higher barrier means lower rate. Tunneling rate depends on the whole shape of the potential and so height itself does not determine rate, and tunneling rate can be higher through a higher potential. A similar effect can occur in superposition of conformations, greater transmission rate through a higher barrier occurring due to the superposition and not due to barrier shape and tunneling. To illustrate this effect without concerns of barrier shape, consider the example of the infinite wall in subsection \ref{infbar}. Tunneling through the infinite wall is zero, but under superposition with a conformation with a delta potential, then, under a certain viewpoint, flux issuing from the  infinite wall can be greater than that from the delta potential, even though, again under a certain viewpoint, the incoming flux to the infinite wall is smaller than that to the delta potential.

To explain what we mean by ``under a certain viewpoint," consider a double slit experiment. One can ask what fraction of incoming particles pass through the upper slit. This is not a well posed question as particles pass through both simultaneously, but one can place a detector after the first slit and consider the rate of detection as an answer of sorts. These are not particles passing through the upper slit but those passing through both slits and detected leaving the upper one. Likewise there is an answer of sorts to the question of how many particles are heading toward the upper slit by placing a detector just before the upper slit. The detection is not the rate of particles heading toward the upper slit but that of particles heading toward both slits and detected in front of the upper one. In other words, one has the \emph{observed} flux issuing from the upper slit and the \emph{observed} flux heading to the upper slit without concerns about the \emph{real} flux which has no meaning in the standard quantum formalism but only in certain hidden variable theories.

In relation to subsection \ref{infbar} we assume we have a projective measurement  with operators \(P_1\) and \(P_2\) in the interaction basis (that of equation (\ref{Schenz})) which in the free basis (that of equation (\ref{Schenzt})) corresponds to \(Q_1\) and \(Q_2\). In the interaction basis the  wave-functions are no longer of  the simple scattering form \(e^{ikq}\) but are superpositions of two of these: \(\Psi=\alpha_1e^{ik_1q}+\alpha_2e^{ik_2q}\). These are still energy eigenstates. The probability current (up to physical constants) \(\displaystyle \frac1{2i}\left(\Psi^* \frac{d\Psi}{dq}-\Psi\frac{d\Psi^*}{dq}\right)\) is
\begin{equation}\label{pc}
  k_1|\alpha_1|^2+k_2|\alpha_2|^2+(k_1+k_2){\rm Re}\,(\alpha_1\alpha_2^*e^{i(k_1-k_2)q}).
\end{equation}
In this expression the first two terms are the familiar fluxes of the two components individually, the third term is the interference term and consists of a flux that varies with position, oscillating between positive and negative. This position dependent flux averages out to zero, so the average flux is given by the first two terms. We use this average flux for comparisons, though other choices, such as maximum or minimum flux, can be made. In the text that follows, ``flux" shall always mean ``average flux". Applying \(Q_1\) and \(Q_2\) to both the incoming and outgoing wave function in the free basis we calculate that the observed  incoming and the observed  outgoing  fluxes for the infinite wall are:
\begin{equation}\label{inwall}
  \alpha k_1|A_1|^2+(1-\alpha)k_2|A_2|^2,
\end{equation}
and
\begin{equation}\label{outwall}
  \alpha k_1|F_1|^2+(1-\alpha)k_2|F_2|^2,
\end{equation}
respectively. The expressions for the observed incoming and observed outgoing fluxes for the delta potential are obtained from these changing \(\alpha\) to \(1-\alpha\).

Using the same numerical values of the parameters as in the example in subsection \ref{infbar}, we have the following figure:

\begin{center}
\includegraphics[scale = 0.3]{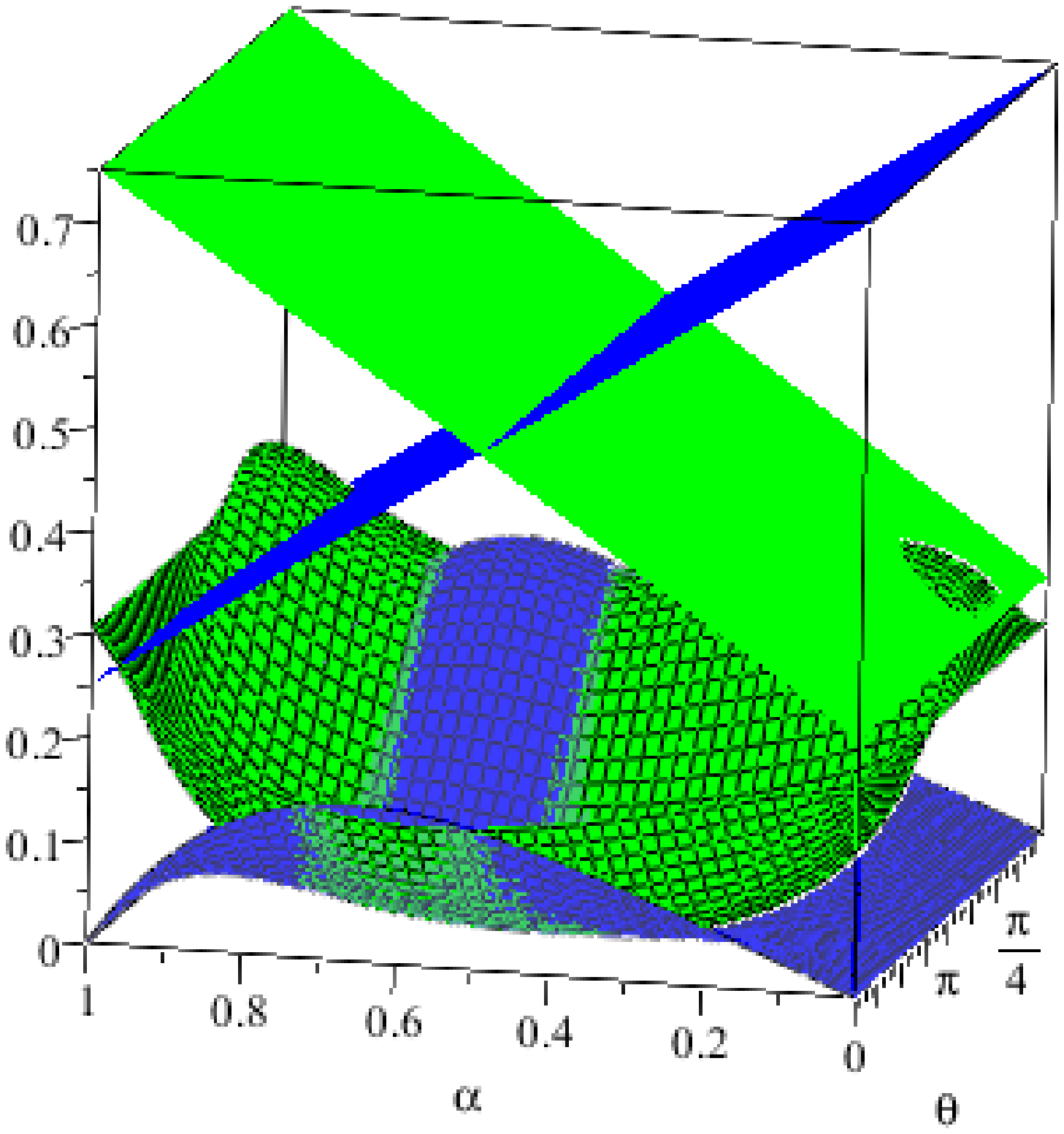}
\end{center}

Here blue color corresponds to the infinite wall, the planar surface is the observed  incoming flux and the curved surface the observed outgoing flux. The green color correspond to the delta potential barrier. One sees here a region where the observed incoming flux to the infinite wall is lower than that to the delta potential while at the same time the observed outgoing flux from the infinite wall is higher than that from the delta potential. Superpositions of conformations can seemingly enhance observed transitions through higher potential barriers even when tunneling cannot account for it. This is illustrated by the  infinite wall case where tunneling is impossible.

\subsection{True bound states}\label{tb}

When one potential is negative and the other of any sign, there can be true bound states depending on the values of the system parameters.

This is no longer an analog of enzyme action but could be the analog of the initial coupling of the two molecules to superimposed conformations. The axis of the \(q\)-variable is no longer the analog of the reaction axis but of the spatial separation of the two molecules. The system parameters for this situation are not to be identified with the system parameters for the transmission processes as the two sets of parameters are related to different phases of the catalytic process.  Interpreting the result in this subsection in relation to real enzymes may be problematic, nevertheless these results show the  existence of bound energy eigenstates with superpositions of conformations, the ``cat states".

Initially we make no assumptions on \(v_1\) and \(v_2\). Equations (\ref{2dogs}) through (\ref{detb}) still hold.

In (\ref{detb}) one can now substitute \(K_i=\sqrt{e_i-E}\) and solve it to obtain the energy levels of the bound states. Since the zero point of energy is arbitrary, we can set \(e_1=0\) without loss of generality. This now gives:
\begin{equation}\label{Elev}
  \sqrt{-E}\sqrt{e_2-E}+(v_2+y(v_1-v_2))\sqrt{-E}+(v_1-y(v_1-v_2))\sqrt{e_2-E}+v_1v_2=0.
\end{equation}
Setting \(u=\sqrt{-E}\) this equation, after some rearrangements becomes
\begin{equation}\label{ulev}
\sqrt{u^2+e_2} = -\frac{(v_2+y(v_1-v_2))u+v_1v_2}{(u+v_1-y(v_1-v_2))}
\end{equation}
We are looking for \emph{positive} solutions of this equation. These occur where the graphs of the functions on both sides intersect for positive abscissa.
We rewrite equation (\ref{ulev}) as
\begin{equation}\label{ulevg}
 \sqrt{u^2+\alpha} = \frac{\beta}{u-\gamma}+\delta
\end{equation}
where \(\alpha=e_2\), \(\beta =y(1-y)(v_1-v_2)^2\), \(\gamma=-(v_1-y(v_1-v_2))\) and \(\delta =-(v_2+y(v_1-v_2))\).

There can only be at most two solutions. Since \(\beta\ge 0\) the graph of the right-hand side consists of two hyperbolic arc that decrease with increasing \(u\). The graph of the left-hand side is a hyperbolic arc that increases. Once the latter intersect one of the arcs of the former, it cannot do so again. Thus there can be no more than two intersection.

When both potentials are negative, there is always at least one solution. This is because then both \(\gamma\) and \(\delta\) are positive. This means that the  first quadrant contains the hyperbolic arc of the right-hand side that opens upwards. The diagonal line in the first quadrant intersects it and since the hyperbolic arc of the left-hand side starts on one of the axes and is asymptotic to the diagonal line, it must intersects it also. An example of  two solution is given by
\(v_1 = -2,\, v_2 = -1,\, y = 1/4,\, \alpha = -1\). The intersection of the two graphs (blue for the right-hand and red for the left-hand side) is shown below:
\begin{center}
\includegraphics[scale = 0.2]{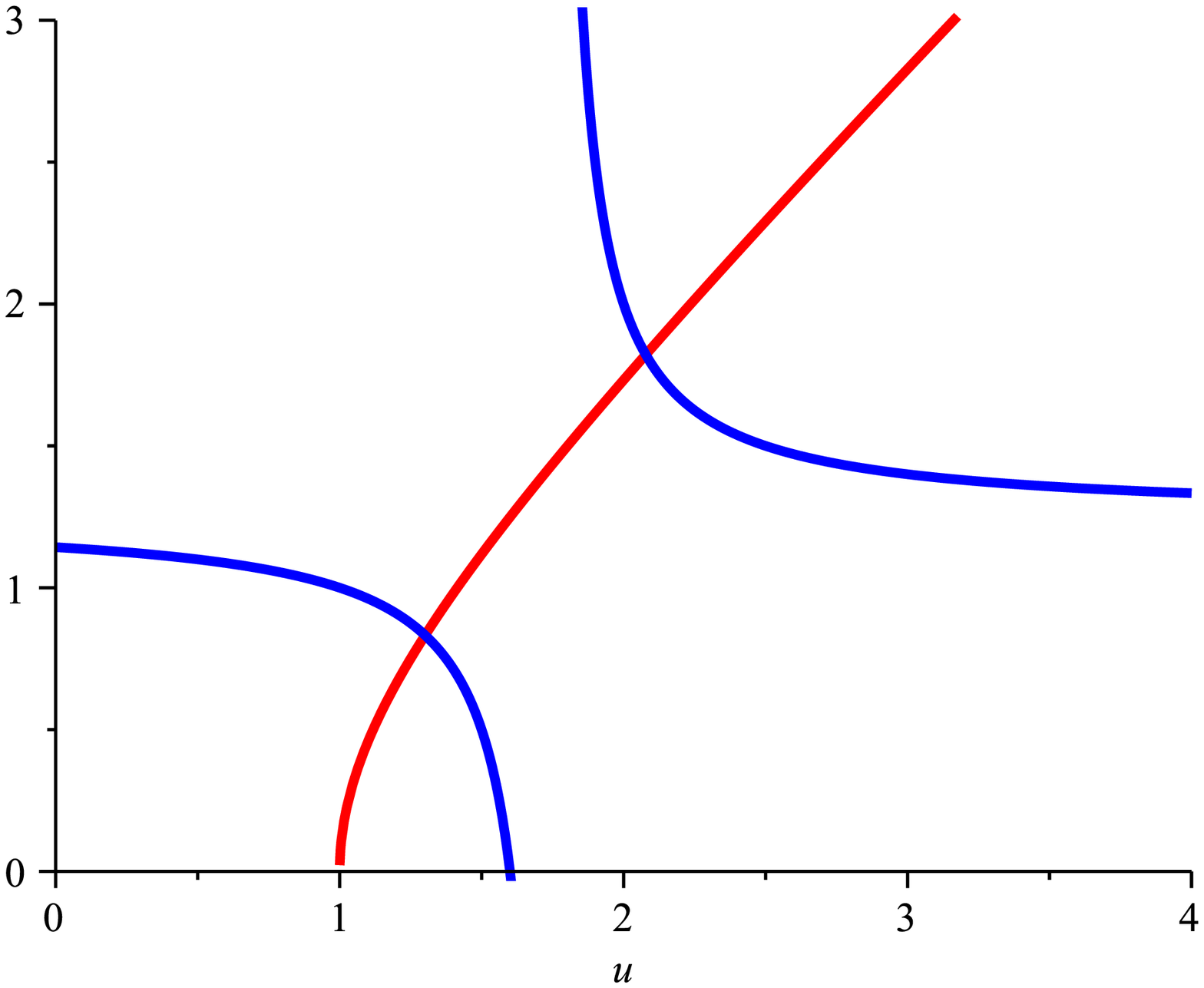}
\end{center}
The two values of energy \(E=-u^2\) are \(-1.6928\) and \(-4.3182\). Changing \(\alpha\) to \(2\) changes the plot to
\begin{center}
\includegraphics[scale = 0.2]{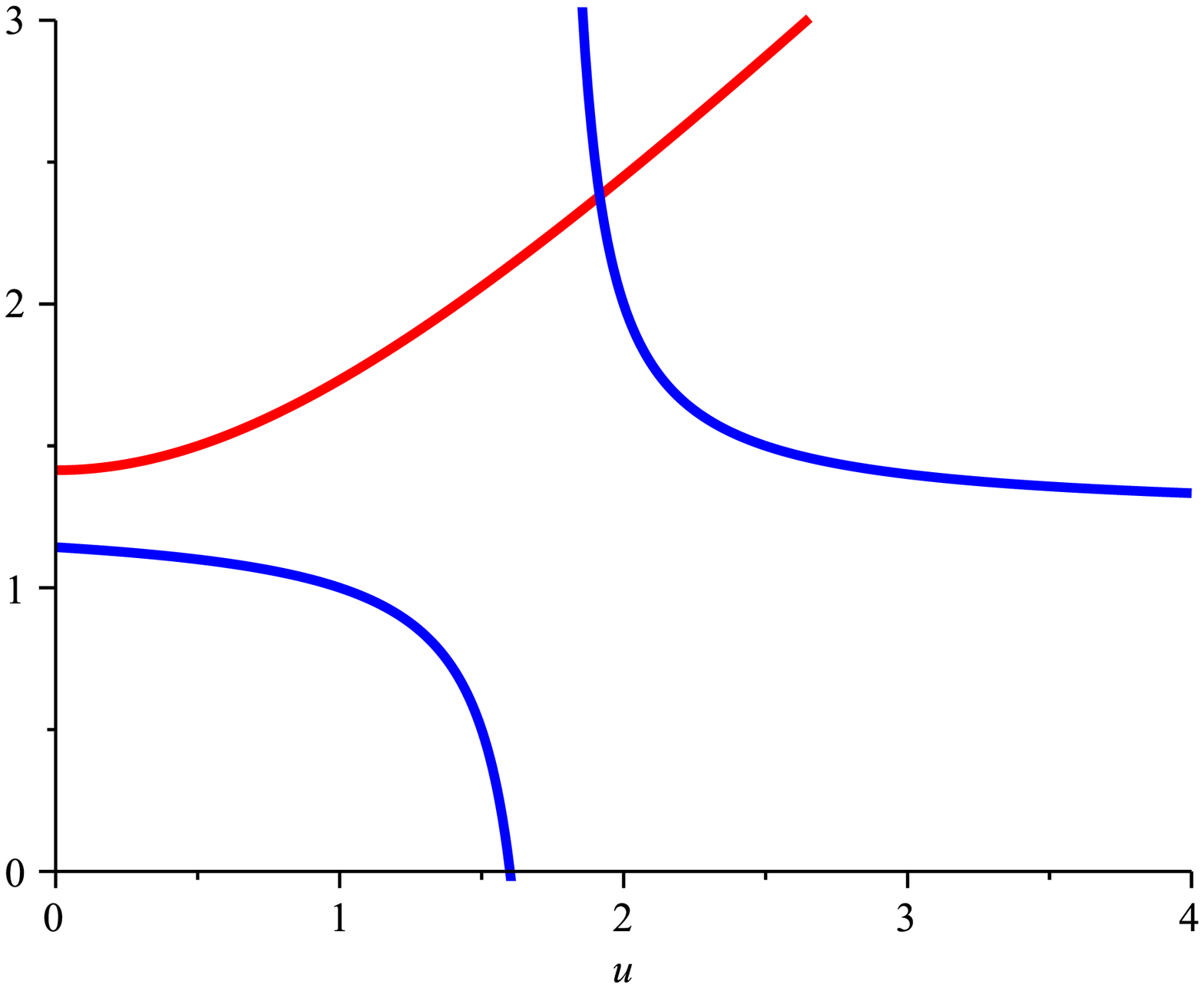}
\end{center}
which has only one point of intersection  corresponding to \(E=-3.6701\).

When only one of the potentials is negative there cannot be two solutions. This follows from a detailed analysis of the toy system whose length makes it fall outside the intended scope of this manuscript which is to present instructive numerical results.

The following examples show cases of one and zero solutions. For \(v_1 = -1,\, v_2 = 1,\, y = 1/4,\, \alpha = 1\) the plot is:
\begin{center}
\includegraphics[scale = 0.2]{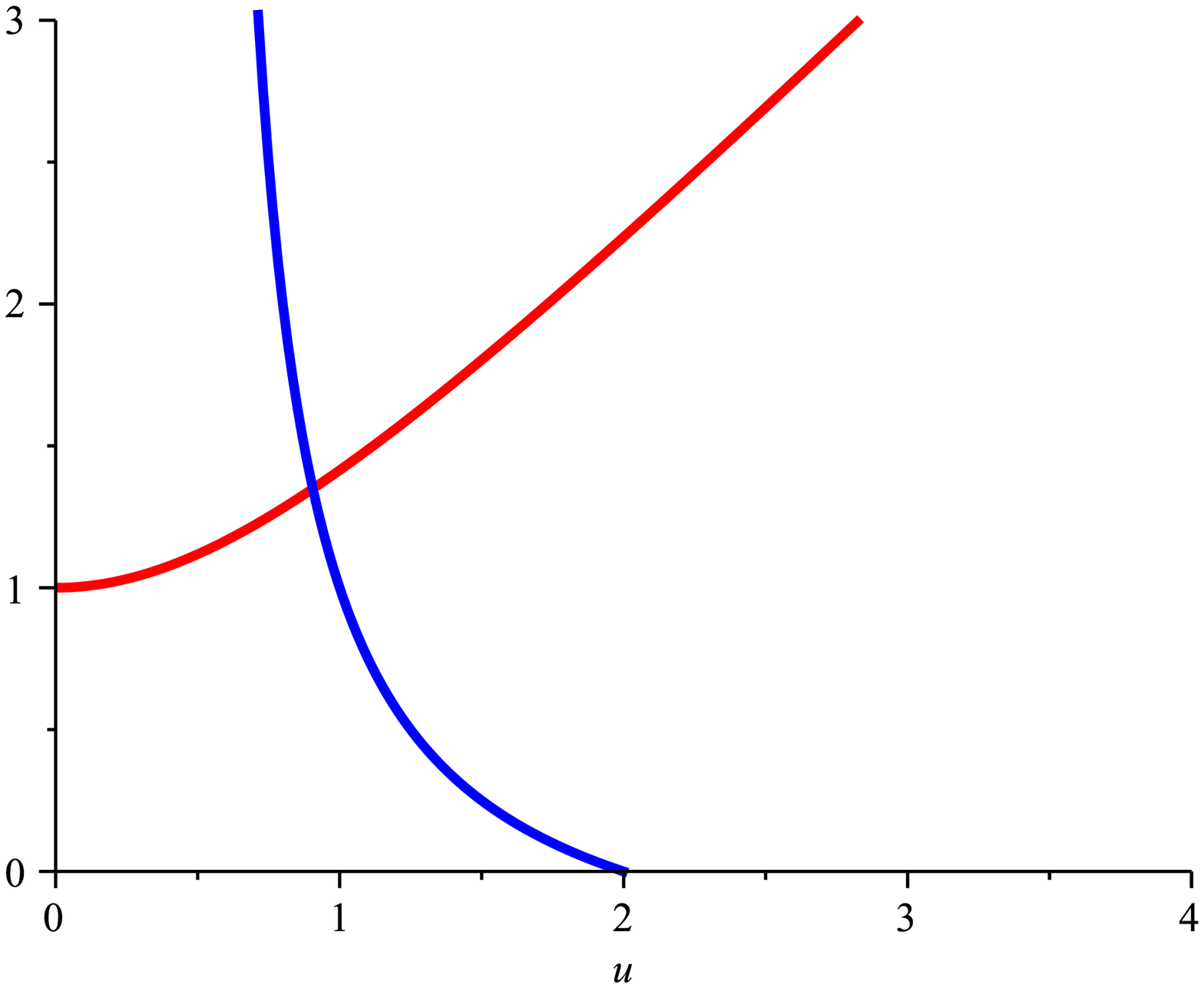}
\end{center}
which shows one solution. Changing \(\alpha\) to \(-5\) changes the plot to
\begin{center}
\includegraphics[scale = 0.2]{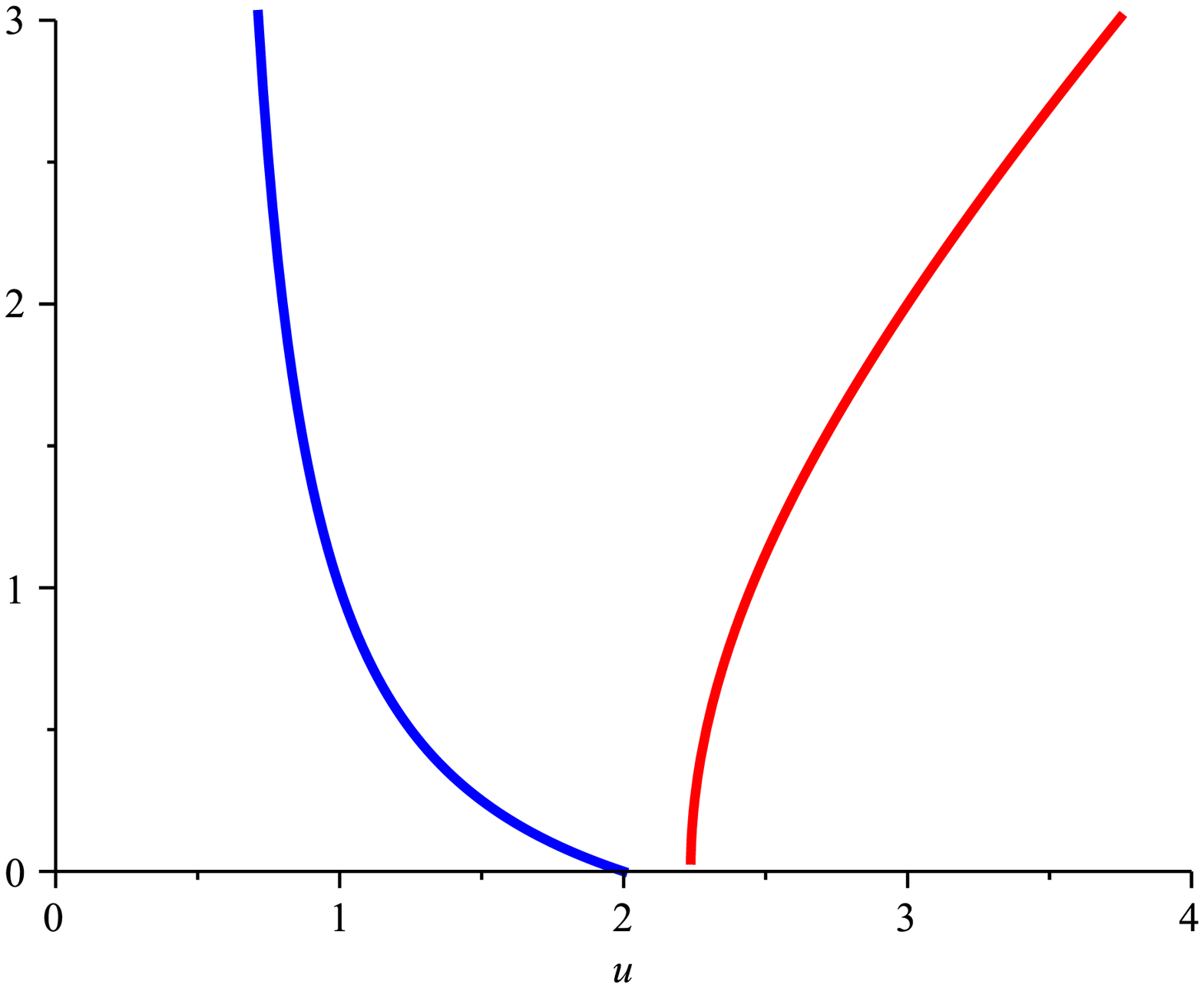}
\end{center}
for which there are no solutions.

A natural question is whether these results suggest that enzyme-substrate binding as cat states is favored over binding in a single conformation. Superficially in the first example one of the cat state energies is \(-4.3182\) while the energies for the delta potentials are \(-1\) and \(-4\). While this is suggestive that a cat state is preferred, it's not a solid conclusion, and better arguments must be devised.

\section*{Acknowledgements}
The author thanks Vasily Ogryzko and Thiago Guerreiro for their interest, helpful ideas, and discussions. Partial financial support was provided by FAPERJ, a government agency of the state of Rio de Janeiro, Brazil.

\end{document}